\documentclass[%
 reprint,
 amsmath,amssymb,
 aps,
]{revtex4-1}

\usepackage{graphicx}
\usepackage{dcolumn}
\usepackage{bm}
\usepackage{amsmath,amssymb,amsfonts,color,fancybox}
\usepackage[breaklinks,colorlinks=true,linkcolor=black,
citecolor=red, urlcolor=blue]{hyperref}

\begin{document}

\preprint{APS/123-QED}

\title{The Dispersion Equation of MGD Waves for non-stationary Plasmas} 

\author{H. P\'erez-de-Tejada}
\author{Eric G\'omez-G\'omez }%
\affiliation{%
 Instituto de Geof\'isica, UNAM, M\'exico, D.F.
}%




\date{\today}


\begin{abstract}
The dispersion equation of MGD plasma waves measured in a reference frame with a relative speed from that where they are generated is derived. The analysis leads further from what is known for waves produced in stationary plasmas in the presence of an imposed magnetic field, and includes terms containing space gradients in the density and speed values in the continuity and momentum equations of the plasma. The dispersion equation that is derived leads to an Alfven wave and magnetogasdynamics wave modes whose distribution with respect the direction of the imposed magnetic field varies with the relative speed of the reference frame. For the particular case in which the speed of the reference frame is zero the dispersion equation of the magnetogasdynamic waves reduces to what is known for stationary plasmas. More complicated waves with different distribution modes with respect to the magnetic field direction are implied from the more general four-order dispersion equation that is derived. A particular case is presented for a non-zero speed value of the reference frame, but the derivation and analysis of the corresponding solutions will be discussed in a forthcoming report.
\end{abstract}
\maketitle


\section{\label{sec:level1}INTRODUCTION}

\begin{figure}[ht]
\begin{center}
\includegraphics{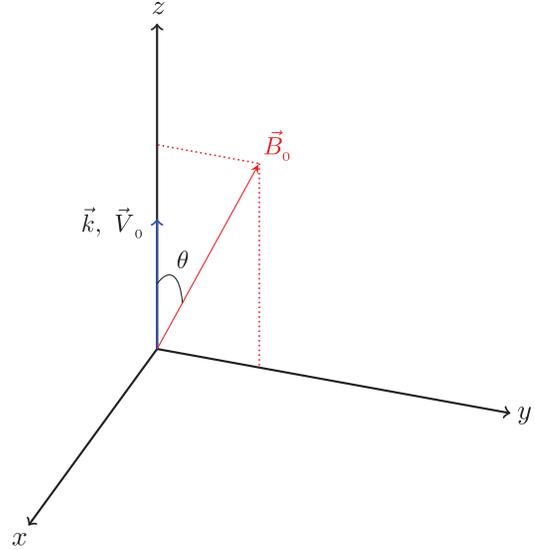}
\caption{\label{Figura:Fig1} Reference frame for MGD waves that propagate along the direction of motion $\vec{V}_{_0}$.}
\end{center}
\end{figure}

The propagation of magnetic and sound waves in perfectly conducting fluids is a common ground in important areas of research such as space physics, Astrophysics and laboratory plasmas. Their analysis and use for stationary plasmas is properly described in various textbooks (\,\cite{Ferraro}, \cite{Hindu}, \cite{Shercliff}, \cite{Hector}\,), and can be further examined by considering peculiar cases in which the plasma is set into motion. The distribution and propagation of such MGD waves can also be viewed as resulting from observations conducted in a reference frame that exhibits a relative speed with respect to the plasma. We will follow a procedure similar to that of electrostatic waves in two streaming instabilities \cite{China}. Within the context of that problem the continuity and the momentum equations of a streaming flow should include terms that account for space gradients in the density and speed of the plasma and that are produced by wave fluctuations different from those expected under stationary plasma conditions. The analysis of hydromagnetic waves that will be presented here also takes account of the conservation of magnetic flux  thus producing a dispersion equation that is suitable for MGD waves that vary as a function of the relative speed between their source frame and that where they are observed. In an analysis similar to that described in \cite{Ferraro} for MGD waves generated under stationary conditions we will consider waves that propagate along the relative motion of the plasma at an angle $\theta$ away from an imposed magnetic field $\vec{B}_{_0}$. We will consider, in particular, the case in which the Alfven speed $V_{_a}$ is larger than the sound speed $V_{_s}$. Such conditions are suitable for the solar wind where usually $V_{_s}\,\approx\,60$ km/s and $V_{_a}\,\approx\,90$ km/s represent most likely values \cite{Gosling}, and where it is possible to consider transport coefficients \cite{CoeficientePerez}. As shown in figure \ref{Figura:Fig1}, the waves are assumed to be directed along the relative speed $\vec{V}_{_0}$ and the magnetic field is in the $yz$ plane; namely, its components are $\bigl(\begin{array} {ccc} 0,& B_{_0}\,\sin\,\theta, & B_{_0}\,\cos\,\theta \end{array}\bigr)$.

\section{MGD WAVES IN A PLASMA WITH RELATIVE MOTION}
The MGD waves will be derived by taking the linear approximation in the following equations:
\begin{itemize}
\item Continuity equation:
\begin{equation}
\dfrac{\,\partial \rho^{\,\prime}}{\partial t} \:+\: \vec{V}_{_0}\cdot\,\bigl(\,\nabla\,\rho^{\,\prime}\,\bigr)\:+\: \rho_{_0}\,\nabla\cdot\vec{V}=0\:. \label{ContinuityEquation}
\end{equation}
\item Momentum equation:
\begin{equation}
\rho_{_0}\,\dfrac{\partial \vec{V}}{\partial t} \: + \: \rho_{_0}\,\vec{V}_{_0}\,\bigl(\,\nabla\cdot\vec{V}\,\bigr) = -V_{_s}^{\,2}\,\nabla \rho^{\,\prime} \:+\: \bigl(\,\nabla\times\vec{b}\:\,\bigr)\times\dfrac{\vec{B}_{_0}}{\mu}, \label{MomentumEquation}
\end{equation}
\end{itemize}

\noindent where we have only considered pressure and magnetic forces. The variation contributions $\vec{b}$, and $\rho^{\,\prime}$, refer to the magnetic field, and density of the plasma. We will follow a derivation similar to that described in \cite{China} for electrostatic waves that arise in two stream flows  with relative motion between them. Accordingly, we will maintain the term \,$\vec{V}_{_0}\cdot\bigl(\,\nabla\,\rho^{\,\prime}\,\bigr)$\, in the continuity equation, and the term \,$\rho_{_0}\,\vec{V}_{_0}\,\bigl(\,\nabla\cdot\vec{V}\,\bigr)$\, in the momentum equation to take into account space gradients in the density and speed of the plasma, and that will lead to MGD waves that were not inferred from the Ferraro and Plumpton analysis \cite{Ferraro}. Both terms combine first order variations of the density and velocity values from waves that will be produced by the relative motion \,$\vec{V}_{_0}$\, with respect to the plasma. Since waves will travel in the $z$-direction all variables depend on $z$ and $t$ only and thus (\ref{ContinuityEquation}) can be written as:
\begin{equation}
\dfrac{\,\partial \rho^{\,\prime}}{\partial t} \:+\: V_{_0}\,\dfrac{\,\partial\rho^{\,\prime}}{\partial z} \:+\: \rho_{_0}\,\dfrac{\,\partial v_{_z}}{\partial z}=0\:. \tag{\ref{ContinuityEquation}a}\label{eq:Continuity}
\end{equation}
At the same time, the coordinate components of (\ref{MomentumEquation}) are
\begin{gather}
\rho_{_0}\,\dfrac{\partial v_{_x}}{\partial t}= \left(\,\dfrac{B_{_0}\cos\,\theta }{\mu}\,\right)\:\dfrac{\partial b_{_x}}{\partial z}\:, \tag{\ref{MomentumEquation}a}\label{eq:MomentumX}      \\[1.6ex]
\rho_{_0}\,\dfrac{\partial v_{_y}}{\partial t}= \left(\,\dfrac{B_{_0}\cos\,\theta }{\mu}\,\right)\:\dfrac{\partial b_{_y}}{\partial z}\:, \tag{\ref{MomentumEquation}b}\label{eq:MomentumY}
\end{gather}
\begin{multline}
\rho_{_0}\,\dfrac{\partial v_{_z}}{\partial t} \:+\: \rho_{_0}\,V_{_0}\,\dfrac{\partial v_{_z}}{\partial z} = -V_{_s}^{\,2}\,\dfrac{\partial \rho^{\,\prime}}{\partial z} \\[1.5ex] - \:\left(\,\dfrac{B_{_0}\,\sin\,\theta}{\mu}\,\right)\:\dfrac{\partial b_{_y}}{\partial z}\:. \tag{\ref{MomentumEquation}c}\label{eq:MomentumZ}\quad
\end{multline}

\noindent where the magnetic field is dictated by the conservation of magnetic flux
\begin{equation}
\dfrac{\partial \vec{b}}{\partial t}= \nabla\times\bigl(\,\vec{V} \, + \, \vec{B}_{_0}  \,\bigr)\:, \label{FluxMagnetic}
\end{equation}

\noindent as it applies to a perfectly conducting gas. In addition:
\begin{equation}
\nabla\cdot\vec{b}=0 \quad\text{then}\quad b_{_z}=0\:.
\end{equation}

\noindent The components of equation (\ref{FluxMagnetic}) are:
\begin{gather}
\dfrac{\partial b_{_x}}{\partial t} = \bigl(\,B_{_0}\,\cos\,\theta\,\bigr)\:\dfrac{\partial v_{_x} }{\partial z}\:, \tag{\ref{FluxMagnetic}a}\label{eq:FluxX}\\[1.6ex]
\dfrac{\partial b_{_y}}{\partial t} = \bigl(\,B_{_0}\,\cos\,\theta\,\bigr)\:\dfrac{\partial v_{_y} }{\partial z}\: - \: \bigl(\,B_{_0}\,\sin\,\theta\,\bigr)\:\dfrac{\partial v_{_z} }{\partial z}\:, \tag{\ref{FluxMagnetic}b}\label{eq:FluxY}
\end{gather}
with no $z$-component for this equation. We will assume harmonic waves of frequency $\frac{\omega}{2\,\pi}$ and wave number $k$ so that the variables vary as \:$\mathbf{e}^{\,i\,(\omega t - kz)}$\:. We obtain:

\begin{gather}
\omega\,\rho^{\,\prime} - V_{_0}\,k\,\rho^{\,\prime} - \rho_{_0}\,k\,v_{_z}=0\:, \tag{\ref{ContinuityEquation}a'}\label{eq:ContinuityArmonic}\\[1.6ex]
\rho_{_0}\,\omega\,v_{_x}=-\left(\,\dfrac{B_{_0}\,\cos\,\theta}{\mu}\,\right)\:k\,b_{_x}\:, \tag{\ref{MomentumEquation}a'}\label{eq:MomentumArmonicX}\\[1.6ex]
\rho_{_0}\,\omega\,v_{_y}=-\left(\,\dfrac{B_{_0}\,\cos\,\theta}{\mu}\,\right)\:k\,b_{_y}\:,\tag{\ref{MomentumEquation}b'}\label{eq:MomentumArmonicY}
\end{gather}
\begin{multline}
\rho_{_0}\,\omega\,v_{_z}\:-\:\rho_{_0}\,V_{_0}\,k\,v_{_z}= \:V_{_s}^{\,2}\,k\,\rho^{\,\prime} \\[1.6ex]
+\: \left(\, \dfrac{B_{_0}\,\sin\,\theta}{\mu} \,\right)\,k\,b_{_y}\:,\quad \tag{\ref{MomentumEquation}c'}\label{eq:MomentumArmonicZ}
\end{multline}

\begin{gather}
\omega\,b_{_x}=-\bigl(\,B_{_0}\,\cos\,\theta\,)\,k\,v_{_x}\:,\tag{\ref{FluxMagnetic}a'}\label{eq:FluxArmonicX}\\[1.6ex]
\omega\,b_{_y}=-\bigl(\,B_{_0}\,\cos\,\theta\,\bigr)\,k\,v_{_y} \,+\, \bigl(\,B_{_0}\sin\,\theta\,\bigr)\,k\,v_{_z}\:. \tag{\ref{FluxMagnetic}b'}\label{eq:FluxArmonicY}
\end{gather}
Equations (\ref{eq:MomentumArmonicX}) and (\ref{eq:FluxArmonicX}) lead together to:
\begin{equation}
\left(\,\dfrac{\omega}{k}\,\right)^{\,2} = \dfrac{B_{_0}^{\,2}\cos\,\theta}{\mu\,\rho_{_0}}\:.
\end{equation}

\noindent At the same time (\ref{eq:MomentumArmonicZ}) can be written as
\begin{equation*}
\bigl(\,\rho_{_0}\,\omega - \rho_{_0}\,V_{_0}\,k\,\bigr)\,v_{_z} - \left(\dfrac{B_{_0}\,\sin\,\theta}{\mu}\right)\,k\,b_{_y}=V_{_s}^{\,2}\,\rho^{\,\prime}\,k\:,
\end{equation*}
where we can replace \,$\rho^{\,\prime}$\, by using (\ref{eq:ContinuityArmonic}):
\begin{equation*}
\bigl(\,\rho_{_0}\,\omega - \rho_{_0}\,V_{_0}\,k \,\bigr)\:v_{_z}\: - \: \left(\,\dfrac{B_{_0}\,\sin\,\theta}{\mu}\,\right)\:k b_{_y}=\dfrac{V_{_s}^{\,2}\,k^{\,2}\,\rho_{_0}\,v_{_z}}{\omega - V_{_0}\,k}\:,
\end{equation*}
and that can also be written as:
\begin{equation*}
-\left(\dfrac{B_{_0}\,\sin\,\theta}{\mu}\right)\,k\,b_{_y}=\left[\, \dfrac{V_{_s}^{\,2}\,k^{\,2}}{\omega-V_{_0}\,k}-\bigl(\,\omega-V_{_0}\,k\,\bigr)\,\right]\rho_{_0}\,v_{_z}
\end{equation*}
or
\begin{equation}
\left(\dfrac{B_{_0}\,\sin\,\theta}{\mu\,\rho_{_0}}\right)\,k\,b_{_y}=\left[\,\bigl(\,\omega - V_{_0}\,k\,\bigr) \,-\, \dfrac{V_{_s}^{\,2}\,k^{\,2}}{\omega - V_{_0}\,k}\right]\:v_{_z}\,. \label{yComponentArmonicModificado}
\end{equation}
\noindent On the other hand, by using (\ref{eq:MomentumArmonicY}) in  (\ref{eq:FluxArmonicY}) we obtain
\begin{equation*}
\omega^{\,2}\,b_{_y} = \left(\,\dfrac{k^{\,2}\,B_{_0}^{\,2}\,\cos^{\,2}\,\theta}{\mu\,\rho_{_0}}\,\right)\:b_{_y} + \bigl(\,B_{_0}\,\sin\,\theta\,\bigr)\omega\,k\,v_{_z}\:,
\end{equation*}
which leads to
\begin{equation}
\left(\,\omega^{\,2}  - \dfrac{k^{\,2}B_{_0}^{\,2}\,\cos^{\,2}\,\theta}{\mu\,\rho_{_0}}\,\right)\:b_{_y}=\bigl(\,B_{_0}\,\sin\,\theta\,\bigr)\,\omega\,k\,v_{_z}\:.\label{yComponentArmonicMagneticModificado}
\end{equation}

\noindent Dividing (\ref{yComponentArmonicModificado}) by (\ref{yComponentArmonicMagneticModificado}) we obtain:
\begin{equation*}
\dfrac{\left(\,\dfrac{B_{_0}\,\sin\,\theta}{\mu\,\rho_{_0}}\,\right)\:k}{\left(\,\omega^{\,2} - \dfrac{k^{\,2}\,B_{_0}^{\,2}\cos^{\,2}\,\theta}{\mu\,\rho_{_0}}\,\right)}\,=\,
\dfrac{\left[\, \bigl(\,\omega - V_{_0}\,k\,\bigr) - \dfrac{V_{_s}^{\,2}\,k^{\,2}}{\bigl(\, \omega - V_{_0}\,k\,\bigr) }\, \right]  }{\bigl(\,B_{_0}\,\sin\,\theta \,\bigr)\,\omega\,k}\:,
\end{equation*}

\noindent where \:$V_{_a}^{\,2}=\dfrac{B_{_0}^{\,2}}{\mu\,\rho_{_0}}$\: is the Alfven speed. This equation can also be written as:
\begin{equation*}
\dfrac{k^{\,2}\, V_{_a}^{\,2}\,\omega\,\sin^{\,2}\,\theta}{\:\omega^{\,2} - k^{\,2}\, V_{_a}^{\,2}\cos^{\,2}\,\theta\:} \,=\, \dfrac{\bigl(\,\omega - V_{_0}\,k\,\bigr)^{\,2} - V_{_s}^{\,2}\,k^{\,2}}{\omega - V_{_0}\,k}\:,
\end{equation*}

\noindent and the dispersion relation is therefore:
\begin{multline}
\omega\, \bigl(\,\omega - V_{_0}\,k\,\bigr)\,\bigl(\,k^{\,2}\,V_{_a}^{\,2}\sin^{\,2}\theta\,\bigr)\:=\: \Bigl[\,\bigl(\,\omega - V_{_0}\,k\,\bigr)^{\,2} - V_{_s}^{\,2}\,k^{\,2}\,\Bigr] \\[1.6ex]
\times\:  \bigl(\:\omega^{\,2} - k^{\,2}\,V_{_a}^{\,2}\,\cos^{\,2}\theta\:\bigr)\:, \label{DispersionEquation}
\end{multline}
\noindent which is a four order equation and thus has four solutions.\\

\noindent As a particular case we take \,$V_{_0}=0$\, and (\ref{DispersionEquation}) reduces to:
\begin{equation*}
\omega^{\,2}\,k^{\,2}\,V_{_a}^{\,2}\sin^{\,2}\theta = \bigl(\, \omega^{\,2} - V_{_s}^{\,2}\,k^{\,2}\,\bigr)\,\bigl(\,\omega^{\,2} - k^{\,2}\, V_{_a}^{\,2}\,\cos^{\,2}\,\theta\,\bigr)
\end{equation*}
or:
\begin{equation}
\left(\dfrac{\,\omega}{\,k}\right)^{\,4} - \bigl(\,V_{_a}^{\,2} + \,V_{_s}^{\,2}\,\bigr)\,\left(\dfrac{\,\omega}{\,k}\right)^{\,2} + \,V_{_a}^{\,2} \,V_{_s}^{\,2}\,\cos^{\,2}\theta =0\:, \label{DispersionEquation-0}
\end{equation}
which is eq. (3.33) of \cite{Ferraro} for MGD waves in stationary plasmas (Under such conditions the four solutions of equation (\ref{DispersionEquation}) reduce to two different modes in equation (\ref{DispersionEquation-0})), \cite{Ferraro}.\\[1.6ex]

On the order hand, if \,$\theta=90^{\circ}$,\, the equation (\ref{DispersionEquation}) becomes:
\begin{equation*}
\omega\, \bigl(\,\omega - V_{_0}\,k\,\bigr)\,(\,k^{\,2}\,V_{_a}^{\,2}\,)\:=\:\omega^{\,2}\,\Bigl[\,\bigl(\,\omega-V_{_0}\,k\,\bigr)^{\,2}- V_{_s}^{\,2}\, k^{\,2}\,\Bigr]\,,
\end{equation*}
or:
\begin{multline}
\left(\dfrac{\,\omega}{\,k}\right)^{\,4} - 2\,V_{_0}\left(\dfrac{\,\omega}{\,k}\right)^{\,3} \,-\, \bigl(\,V_{_a}^{\,2} + V_{_s}^{\,2}- V_{_0}^{\,2}\,\bigr)\left(\dfrac{\,\omega}{\,k}\right)^{\,2} \\[1.6ex]
 +\: V_{_a}^{\,2}\,V_{_0}\,\left(\dfrac{\,\omega}{\,k}\right)=0\:.\:\label{DispersionEquation90}
\end{multline}
\begin{figure}
\includegraphics{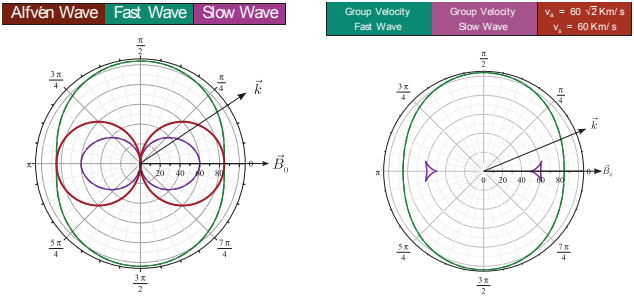}
\caption{\label{Figura:Fig2} (left panel) Phase speed values obtained from equation (\ref{DispersionEquation-0}) for an Alfven wave (red curve) and for the slow (violet curve) and fast (green curve) wave solutions when $V_{_0}=0$ km/s. (right panel) Wave fronts obtained from the group speed values corresponding to the slow (inner curve) and the fast mode (outer curve) of the left panel and that are in agreement with those presented by \cite{Ferraro}.}
\end{figure}
and, at the same time, if $\theta=0^{\circ}$, (\ref{DispersionEquation}) is:
\begin{equation*}
k^{\,2}V_{_a}^{\,2}\, + \,\omega^{\,2}=0 \quad \text{and}\quad \bigl(\,\omega - V_{_0}\,k\,\bigr)^{\,2} - V_{_s}^{\,2}\,k^{\,2}=0\:,
\end{equation*}
leading to: $\left(\dfrac{\,\omega}{\,k}\right) = \pm\,V_{_a}$ and $\left(\dfrac{\,\omega}{\,k}\right) = \pm\,\bigl(\,V_{_0} + V_{_s}\,\bigr)\:,$\\

\noindent which is indicative of an Alfven and sonic waves moving along the magnetic field direction. The phase speed pattern and the shape of a wave front derived under stationary conditions in equation (\ref{DispersionEquation-0}) are shown in figure \ref{Figura:Fig2}, and agree with those derived by \cite{Ferraro} and other authors.\\


The numerical solutions of the general dispersion equation (\ref{DispersionEquation}) for non-zero $V_{_0}$ speed values are very different from these figures and will be presented in an extended report \cite{Tesis}. The different phase velocity plots and wave fronts that are produced by varying the relative speed between the source frame and the reference frame where the waves are observed lead to patterns that do not imply a frequency shift as it occurs in the Doppler effect.\\

\begin{figure}
\includegraphics{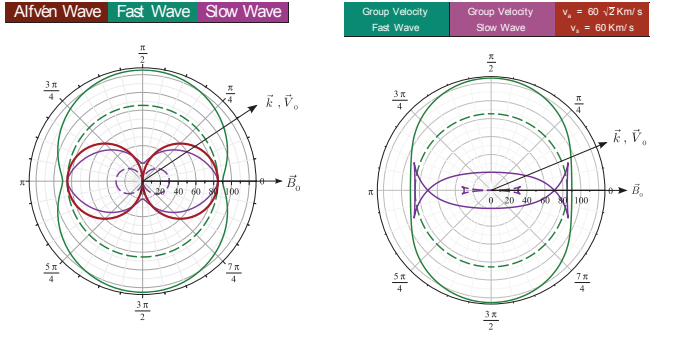}
\caption{\label{Figura:Fig3} (left panel) Phase speed values obtained from equation (\ref{DispersionEquation}) for an Alfven wave (red curve) and for the slow (two violet curves) and fast (two green curves) MGD waves obtained when $V_{_0} = 30$ km/s. (right panel) Wave fronts obtained from the group speed values corresponding to the slow mode waves (violet curves) and to the fast mode waves (green curves) of the left panel.}
\end{figure}

An example of those solutions is presented in figure \ref{Figura:Fig3} for the case in which $V_{_0}=30$ km/s. While the shape of an Alfven wave (red curve) in the phase speed (left) panel  remains unchanged with respect to that in figure \ref{Figura:Fig2} it is notable that  the slow (violet curve) and the fast (green curve) waves of figure \ref{Figura:Fig2}  have split into two different solutions for each case. In particular, one of the two slow (violet) solutions shows decreased values (dashed curve) and the other larger values (full curve) than those indicated in figure \ref{Figura:Fig2} for stationary plasmas. At the same time, the two fast (green) solutions lead to deformed curves where one of them exhibits smaller values than those in figure \ref{Figura:Fig2} at all angles around the magnetic field direction (dashed curve) and the other a dented profile with speed values that are smaller at and in the vicinity of the magnetic field direction.\\

In a similar way as for stationary plasmas in figure \ref{Figura:Fig2}  it is possible to derive wave fronts that the phase speed plots in the left panel of figure \ref{Figura:Fig3} produce on the group speed values. These are reproduced in the right panel of figure \ref{Figura:Fig3} to show very distinct patterns from that in figure \ref{Figura:Fig2}. The small cusp in the vicinity of the magnetic field produced by the slow wave (violet curve) in the left panel of figure \ref{Figura:Fig2} has been replaced by two different (violet) curve shapes. One of them is a narrow curve nearly oriented along the magnetic field direction and the other provides larger size cusps in the vicinity of the magnetic field direction with an added bulge that extends at other angle values. By the same token the fast (green) curve in the right side panel of figure \ref{Figura:Fig2} has also been replaced by two different curves. While both of them show smaller values than those in the curve of figure \ref{Figura:Fig2} there is a larger tendency in one of them for this effect at large angles from the magnetic field direction, while the other keeps nearly the same values.


\bibliographystyle{plain}
\bibliography{BIBLIOPaperDispersion}
\nocite{*}

\end{document}